\documentclass[letterpaper]{article} 
\usepackage{aaai25}  
\usepackage{times}  
\usepackage{helvet}  
\usepackage{courier}  
\usepackage[hyphens]{url}  
\usepackage{graphicx} 
\usepackage{natbib}  
\usepackage{caption} 
\usepackage{algorithm}
\usepackage{algorithmic}

\usepackage{xcolor}
\newcommand{\answerYes}[1]{\textcolor{blue}{#1}} 
\newcommand{\answerNo}[1]{\textcolor{teal}{#1}} 
\newcommand{\answerNA}[1]{\textcolor{gray}{#1}} 
 
\usepackage{newfloat}
\usepackage{listings}
\DeclareCaptionStyle{ruled}{labelfont=normalfont,labelsep=colon,strut=off} 
\floatstyle{ruled}
\newfloat{listing}{tb}{lst}{}
\floatname{listing}{Listing}
\title{UKTwitNewsCor: A Dataset of Online Local News Articles\\for the Study of Local News Provision}
\author {
    Simona Bisiani\textsuperscript{\rm 1},
    Agnes Gulyas\textsuperscript{\rm 2},
    John Wihbey\textsuperscript{\rm 3},
    Bahareh Heravi\textsuperscript{\rm 1}
}
\affiliations {
    \textsuperscript{\rm 1}Surrey Institute for People-Centred AI, University of Surrey\\
    \textsuperscript{\rm 2}Canterbury Christ Church University\\
    \textsuperscript{\rm 3}Northeastern University\\
    s.bisiani@surrey.ac.uk
}

\begin{document}

\maketitle

\begin{abstract}
In this paper, we present \texttt{UKTwitNewsCor}, a comprehensive dataset for understanding the content production, dissemination, and audience engagement dynamics of online local media in the UK. It comprises over 2.5 million online news articles published between January 2020 and December 2022 from 360 local outlets. The corpus represents all articles shared on Twitter by the social media accounts of these outlets. We augment the dataset by incorporating social media performance metrics for the articles at the tweet-level. We further augment the dataset by creating metadata about content duplication across domains. Alongside the article dataset, we supply three additional datasets: a directory of local media web domains, one of UK Local Authority Districts, and one of digital local media providers, providing statistics on the coverage scope of \texttt{UKTwitNewsCor}. Our contributions enable comprehensive, longitudinal analysis of UK local media, news trends, and content diversity across multiple platforms and geographic areas. In this paper, we describe the data collection methodology, assess the dataset geographic and media ownership diversity, and outline how researchers, policymakers, and industry stakeholders can leverage \texttt{UKTwitNewsCor} to advance the study of local media.
\end{abstract}

\begin{links}
     \link{Dataset}{https://doi.org/10.7910/DVN/R5XTEO}
 \end{links}

\section{Introduction}
Local media is essential to democracy, but decades of declining advertising revenue, falling newspaper circulation, and challenges in monetising digital content have eroded local news provision, both in the UK and globally \cite{cairncross_cairncross_2019}. This local news crisis has led to the closure of countless news media outlets \cite{franklin_downgrading_2006, bisiani_uncovering_2023}, significant consolidation within the industry \cite{the_media_reform_coalition_who_2023}, and perceived decline in the quality of local journalism. The latter encompasses various news production phenomena, including content homogenisation, reduction in coverage of local news, and lack of in-depth reporting \cite{fenton_deregulation_2011, franklin_downgrading_2006, davies_flat_2011, johnston_churnalism_2017}.\newline
These conditions have fostered governmental debate into the sustainability of the local media sector and the supply of public interest journalism \cite{cairncross_cairncross_2019, ofcom_review_2024}. In this context, empirical assessments of local news content can provide evidence on how news provision differs across providers and which communities are underserved. Such granular understanding of local news provision can be leveraged to design interventions aimed at sustaining the sector \cite{moore_local_2024}, particularly in terms of where interventions are most urgently needed and how to distribute resources across providers. Yet, empirical studies of news content in the UK remain scarce \cite{broersma_exploring_2018}, potentially due to challenges inherent with collecting local news data \cite{moore_local_2024}.\newline
We aim to contribute in filling this gap by creating \texttt{UKTwitNewsCor}, a dataset comprising over 2.5 million online local news articles from 360 commercial and independent UK news websites, spanning a 36-month period (January 1st, 2020, to December 31st, 2022). Using a seed directory of X (then Twitter) handles of local media brands, we collected tweets from these outlets via the Twitter Academic API and extracted embedded articles originating from the outlet's news website. We enhanced this collection by integrating metadata on the media provider, the Local Authority District(s) targeted by the media outlet which published the article, and social media engagement metrics at the tweet level. We additionally generate metadata on whether the article is duplicated within the corpus or whether it highly resembles any other article within the corpus. We contribute three additional datasets: a directory of UK local news domains, a directory of Local Authority Districts (LADs) with media provision statistics, and a publisher-level directory with market-level statistics. This paper outlines the methodology employed to construct the dataset, assesses its inferential potential with a focus on domain, ownership, and geographic coverage, and highlights key research opportunities that it unlocks. We summarise here \texttt{UKTwitNewsCor}'s contributions:
\begin{itemize}
    \item Offering a UK-specific focus, addressing a gap in local news datasets tailored to the UK market.
    \item Providing extensive coverage of local media providers and administrative geographies, including 47\% of local media web domains active at the time of data collection, covering 94\% of UK Local Authority Districts (LADs).
    \item Supporting diverse and representative sampling strategies (e.g., stratified random or systematic), enhancing the reliability and validity of content analyses. The time frame allows for the identification of trends in local media coverage, providing insight into shifting journalistic priorities, audience engagement, and content production.
    \item The time frame captures the enduring effects of the Covid-19 pandemic, including localised lockdowns, the rollout of vaccines, and the associated socioeconomic challenges, particularly in relation to public health, local businesses, and regional economies. It also spans the post-Brexit transition, highlighting its impact on local industries, trade, and cross-border relations, especially in Northern Ireland. Politically, it covers the outcomes of the 2019 General Election and the subsequent regional responses, as well as the cost-of-living crisis, which had direct consequences for local residents’ daily lives. The time frame allows longitudinal analyses of the long-term impact of these events on local media, reducing the risk of short-term biases that might arise from focusing on a single event or year.
    \item The dataset is suited to exploring production strategies, information flows, and the impact of algorithmic control on digital local journalism amidst increasing dependency on platforms to reach audiences.
    \item Facilitating robust analyses of content duplication practices resulting from media ownership consolidation, as well as geographic and organisational differences in news provision.
\end{itemize}

\section{Motivation}
Information about the workings of local public authorities and services is vital for democracy. Citizens require access to basic local information to engage meaningfully in their communities and navigate everyday life \cite{philip_m_napoli_matthew_weber_kathleen_mccollough_approach_2018, friedland_review_2012}. Communities with weaker media infrastructure tend to experience poorer public resource management, higher corruption, and lower political engagement \cite{gao_financing_2020, hayes_decline_2018}. Worryingly, decades of instability in the UK local media sector, marked by widespread redundancies, closures, and media ownership consolidation, have significantly diminished the quality and quantity of local news reporting. The closure of hundreds local news outlets in the UK since the turn of the century \cite{hunter_uk_2024} has generated geographic disparities in access to local news \cite{gulyas_local_2021, bisiani_uk_2024}. Meanwhile, the local news market in the UK has become increasingly consolidated, with 57\% of print and digital titles owned by the three largest publishers \cite{the_media_reform_coalition_who_2023}. Pursuing economies of scale, these providers have consolidated distinct local newspapers under unitary regional news websites \cite{moore_local_2024}, reducing the number of newsrooms \cite{tobitt_reach_2021} and reporters \cite{ponsford_colossal_2024}. Said consolidation has led to the practice of recycling news content across sister publications and a loss of in-depth reporting, undermining the provision of genuine local reporting and the watchdog role of local media \cite{johnston_churnalism_2017, phillips_faster_2011}. Compelled by platform dependency and the imperatives of metrics, media publishers increasingly align their content strategies with algorithmic logics that privilege engagement over substance \cite{wihbey_social_2019}. However, social media algorithms incentivise the production of sensational and nationally-focused content, often to the detriment of in-depth local journalism \cite{toff_is_nodate}. This dynamic not only undermines the quality of local news coverage but also risks diminishing audience trust in the long term \cite{kormelink_what_2018}.
The increasing role of social media platforms in digital gatekeeping \cite{ofcom_review_2023, bro_digital_2024} underscores the urgency of studying how engagement-driven strategies affect local news visibility and editorial control. Meanwhile, the precise variations in the quality and characteristics of local news coverage across different regions, media providers, and social media platforms remain unclear. This knowledge gap hampers efforts to assess local media’s effectiveness in meeting communities' information needs and limits policymakers' ability to address disparities in news access \cite{moore_local_2024}. Therefore, a promising research avenue consists of evaluating local media content, identifying which communities are underserved and which publishers are fulfilling their public service obligations. Such evaluations can be used to design targeted interventions aimed at improving resource allocation, supporting local journalism, and ensuring equitable news coverage for underserved communities \cite{ofcom_review_2024, communications_and_digital_committee_future_2024}.

\subsection{The Problem of Scale: Addressing Gaps in Local Media Content Analysis}
Empirical content analyses of local media have historically been synonymous with expensive and difficult data collection and analysis \cite{moore_local_2024}. Unlike national or international news outlets, local media are poorly available in news archives and lack of Access Point Interfaces (APIs) for structured and effective data collection. Meanwhile, collecting data directly from publishers is resource-intensive due to the unstructured nature of news websites \cite{riffe_analyzing_2019}. This has led much local media research in the UK to seek to evaluate changes in news quality through newsrooms or audience research \cite{mathews_life_2022, barclay_local_2022, broersma_exploring_2018}. Scholars have long voiced the concern that our knowledge of local news outputs is often impressionistic \cite{bromley_making_2005}, calling for more data-driven approaches \cite{cox_city_1973, griffin_city_2002, franklin_what_1991}. Past content analyses of local media conducted in the UK have contributed unique and useful evidence on the characteristics of local media coverage \cite{bromley_making_2005}, the shift from local to national coverage \cite{franklin_downgrading_2006}, the role of hyperlocal media \cite{williams_value_2015}, and the performance of local news during the 2019 election campaign \cite{moore_local_2024}. Yet, their generalisability is constrained by limited scope across either outlet sample size, article sample size, or time. Bromley \shortcite{bromley_making_2005}, who due to the large diversity of the sector focused on a purposive sample of newspapers published in south-east Wales, recognised that working on a small sample limited his ability to establish causal relationships and generalise to the wider context of local news across the country. The most extensive audit of local media content to date was conducted by Moore and Ramsay \shortcite{moore_local_2024}, who analysed election coverage across 76,766 articles published within one December week in 2019 and representative of around 9 in 10 digital local media brands. While insightful, a purposive sample of articles representing a single week of coverage preceding a general election may limit the generalisability of the findings, making it difficult to extrapolate trends across the rest of the year \cite{long_obtaining_2005}. These issues constitute 'the problem of scale', described as the challenge of using any specific example to say something larger about journalism \cite{carlson_temporal_2019}. Content analysis, defined as making replicable and valid inferences from texts to their contexts \cite{krippendorff_content_2019}, requires careful attention to sampling strategies to ensure both reliability and validity. Sampling local media content presents unique challenges due to variations in outlet size, geography, and socio-political context. Purposive and convenience sampling are simple approaches but they limit the generalisability of findings; more sophisticated approaches, such as random and stratified random sampling, improve representativeness by addressing population disparities \cite{long_obtaining_2005}. Constructed week sampling is one common approach which has proven effective for capturing temporal variations in news coverage \cite{lacy_sample_2001}. In this context, large-scale datasets allow to design sophisticated sampling techniques while also supporting automated analyses methods requiring larger datasets, such as machine learning \cite{hox_computational_2017}.

\subsection{Leveraging Computational Approaches and Large-Scale Datasets for Local News Analysis}
Advances in computational methods have revolutionised journalism research, enabling large-scale data collection and analysis that reduce cost and increase efficiency \cite{zamith_content_2015, chang_understanding_2014}. Automated tools, such as web crawling and APIs, are critical for collecting digital content efficiently. Twitter’s (now X) Academic API exemplified this potential, allowing researchers to access extensive data until its discontinuation in 2023, which disrupted many research projects \cite{murtfeldt_rip_2024, ledford_researchers_2023}. Despite these setbacks, platforms like X remain integral to understanding the dynamics of local news dissemination.\newline
This study capitalised on the final availability of Twitter’s Academic API to collect extensive data on the social media activity of UK local news providers and link this data to news outputs. This paper details the methodology for constructing \texttt{UKTwitNewsCor}, explores its potential for robust inference, and identifies potential use cases and research priorities.

\section{Methodology}
\begin{figure*} 
	\includegraphics[width=\linewidth]{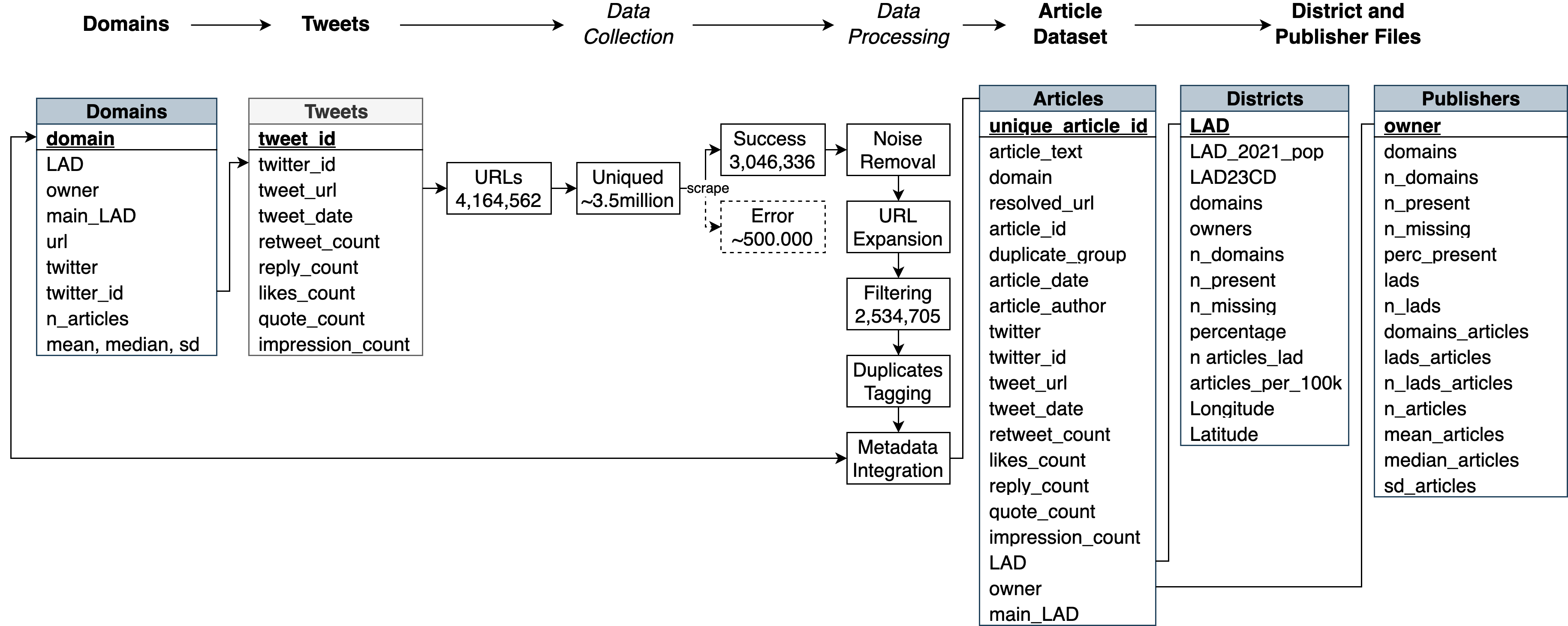}
	\caption{Datasets and Procedure Overview} 
	\label{fig:sys_overview}
\end{figure*}
\subsection{Domains Collection}
Defining "local media" poses theoretical challenges, complicating its operationalisation in research contexts \cite{gulyas_introduction_2020}. In this study, we adopt a definition consistent with prior literature, describing local media as any media entity with a clearly identifiable geographic target audience at a subnational level \cite{gulyas_local_2021}. This definition encompasses a wide range of outlets, from major regional media brands to hyperlocal parish magazines. In the UK, local media manifests in various forms, including newspapers, radio stations, television channels, and online platforms, each catering to distinct population segments \cite{ofcom_review_2024}.\newline 
Online local media is similarly diverse. Moore and Ramsay \shortcite{moore_local_2024} classify online local media in the UK into four types: (1) websites tied to print newspaper brands, (2) regional aggregation hubs owned by major publishers like Reach PLC and Newsquest, (3) digital-only platforms from traditional publishers, and (4) independent hyperlocal news sites focused on community affairs. In addition to these, some local media operate solely on social media platforms, such as Facebook pages or WhatsApp groups, which are increasingly used to disseminate local news and engage with communities \cite{ofcom_review_2024}. For this study, we utilised a directory of 1,070 print and digital local news outlets compiled by Bisiani and Heravi \shortcite{bisiani_uncovering_2023} as our primary data source. The directory was constructed by integrating data from publicly available sources, such as circulation auditors’ member directories, and was refined through manual verification and research. We used an enhanced version of the openly released dataset \cite{bisiani_print_2023} which we obtained by the authors. Differently from the open version, this edition contained additional information about whether an outlet was print or digital, as well as the media provider of each outlet. The directory excludes local media with solely a social media presence, as these are challenging to capture at scale \cite{bisiani_uk_2024}. The directory accounts for print editions of online media separately, distinguishing titles that are unique in print but consolidated online under a shared domain. For example, Solihull Observer, Arden Observer, and Shirley Observer are recognised as distinct print brands in the directory, but they all operate under the unified online domain solihullobserver.co.uk. As we are collecting data from the online domains, we crafted a unique list of web domains from the directory, resulting in 770 domains. For each domain, the original directory provided the publisher and Local Authority District (LAD) of coverage. LADs represent the standard level of administrative geography used when assessing local news provision, and there are 361 across the UK \cite{bisiani_uk_2024, bisiani_uncovering_2023}.

\subsection{Tweets Collection}
To extract Twitter handles for each domain derived from the directory of outlets, the process involved programmatically accessing each URL and retrieving the corresponding webpage content. The HTML of each page was parsed to identify links containing "twitter.com," using a search pattern designed to locate Twitter profile URLs. This procedure allowed for the identification of 450 Twitter handles. These handles were manually checked to verify their relevance. Beyond the collected handles, some Twitter accounts may have been missed due to variations in website structures, inaccessible pages, or alternative methods of embedding social media links. Further attempts to expand the search were limited by time constraints, as we submitted an application for access to X's (then-Twitter) Academic API and, upon approval, had a very short window to collect data before the API was discontinued in March 2023. This data collection was conducted in January and February of 2023 using an R script built with the R library AcademicTwitteR \cite{BarrieHo2021}. We retrieved all tweets posted by the identified handles between 1st January 2020 and 31st December 2022. This resulted in a dataset comprising 4,721,846 tweets, enriched with metadata including like counts, retweet counts, and quote counts.

\subsection{Articles Collection}
We filtered the tweets to retain those that contained a URL, yielding 4,164,562 URLs, which went down to approximately 3.5 million URLs after removing duplicates. As not all handles had shared tweets with URLs, the number of handles went down to 404. Due to the large volume of URLs, we used a parallelised Python script leveraging the \textit{newspaper3k} \cite{ou2013newspaper3k} library to extract article content and metadata from the URLs. Similar to Khanom et al. \shortcite{khanom_news_2023}, we encountered challenges due to paywalls, site design, and inconsistent metadata, causing failure to retrieve articles from approximately 500,000 URLs. For 26 handles, this implied a 100\% retrieval failure. The remaining 378 handles collectively retrieved 3,046,336 articles. We proceeded to inspect and refine the article dataset, through the following steps:
\begin{itemize}
\item \textbf{Noise Removal} -
We identified 556,259 articles containing standardised noise patterns (e.g., newsletter promotions) and cleaned them using sentence tokenisation and pattern matching.
\item \textbf{URL Resolution} -
The URLs were sometimes compressed by URL shorteners (e.g., bit.ly). We used a script to resolve redirects and retrieve final URLs that returned HTTP 200 codes.
\item \textbf{Domain Filtering} -
We did not expect all URLs present in the Twitter feeds of local media accounts to be links to news articles from the publisher. Some were links to governmental or local council websites, or to the BBC. For each handle, we only retained articles that matched the associated domain and discarded the remainder. The final dataset contains 2,534,705 articles. 
\item \textbf{Duplicate Content Tagging} -
We noticed the presence of identical and nearly-identical articles across domains. To facilitate future research focused on content re-usage practices, we created metadata that identifies groups of highly similar or identical articles published across different publications. Due to the size of the data, we applied Locality-Sensitive Hashing (LSH) \cite{jafari_survey_2021} to efficiently detect duplicates. Articles were preprocessed and hashed with MinHash \cite{broder_resemblance_1997}, compared using Jaccard similarity \cite{niwattanakul2013using}, and clustered if their similarity score exceeded 0.7 (see Appendix A for more detail about this procedure). Each article was then tagged with its group identifier. 
\item \textbf{Metadata Integration} -
We used the directory by Bisiani and Heravi \shortcite{bisiani_uncovering_2023} to attach domain-level metadata to each article, including information about the domain, the publisher, and the Local Authority District(s) of coverage.
\end{itemize}

\section{Dataset Overview}
In the articles dataset, each row corresponds to a unique web url containing an article, and a key variable \textit{unique\_article\_id} has been created as an identifier for each article (a complete list of variable names and descriptions can be found in Appendix B). Due to the nature of the data collection, the following conditions apply: (1) An article which has been shared twice by the same account or by different accounts on Twitter will be displayed as one row. Twitter-level variables, in the format of list-columns, contain information regarding the individual twitter events in which any given article was shared, and provide the singular tweet-level metadata for each event (replies, retweets, posting date, \dots). (2) Articles which have identical siblings across different domains are kept separate and are denoted by the variable \textit{article\_id}. (3) Articles which were detected as nearly identical duplicates are kept separate, regardless of whether they were published by the same domain or different ones, and are denoted by the variable \textit{duplicate\_group}, containing the list of applicable \textit{article\_id}s.

\subsection{Coverage Evaluation and Additional Datasets Creation}
To evaluate \texttt{UKTwitNewsCor}, we calculated the number of domains under each provider and within each district within our dataset. We computed the statistical distributions of domains across providers and districts and compared them, by means of descriptive statistics and density curves, to the distributions within the complete list of domains, including domains missing from our dataset. We report on these findings in section \ref{results}. At this point, we produced two additional files to document coverage extents across districts and providers and facilitate future sampling strategies and inferential assessments: (1) a \textit{LAD dataset} where each row corresponds to a unique Local Authority District, and (2) a \textit{publisher dataset} where each row represents a provider. In the district-level table, we incorporated supplementary population size information from the Office for National Statistics \shortcite{ONS_Local_Authority_Districts_2024}, which we used to generate an estimate of the number of articles per 100,000 people.

\subsection{Distribution, FAIR Principles, and Use Guidance}
To facilitate research, we provide the data in two formats: as four separate CSV files (articles.csv, domains.csv, districts.csv, and publishers.csv) or as an SQLite3 database with the four respective tables (see Figure \ref{fig:sys_overview} for an overview of how the datasets are related). We provide a list and description of the variables in Appendix B (Tables \ref{tab:articles_description}, \ref{tab:domains_description}, \ref{tab:districts_description}, \ref{tab:publishers_description}). The \texttt{UKTwitNewsCor} dataset is publicly available in the Harvard Dataverse repository under a CC BY-NC 4.0 License. To ensure reproducibility and provenance, the dataset described in this paper will remain static. However, the authors commit to maintaining the dataset by providing updates regarding its use, addressing any issues that arise, and offering ongoing support to users. \texttt{UKTwitNewsCor} adheres to the FAIR principles. The dataset is Findable, as it is stored in the Harvard Dataverse, with a globally unique and persistent identifier. It is Accessible and Interoperable, as the data is retrievable through the repository's graphical user interface (GUI) and provided in multiple standard formats, CSV and SQLite3, which are compatible with a wide range of tools and programming languages. The dataset is Reusable, as it is distributed under the CC BY-NC 4.0 license, includes comprehensive documentation, and retains original URLs for each article, maintaining clear provenance. The data provided is meant to be used for responsible research on local media in the UK. This includes recognition of usage of copyrighted work by news media organisation, which may not be used to develop commercial applications leveraging their articles without their consent.

\begin{figure*}[!ht] 
	\includegraphics[width=0.97\linewidth]{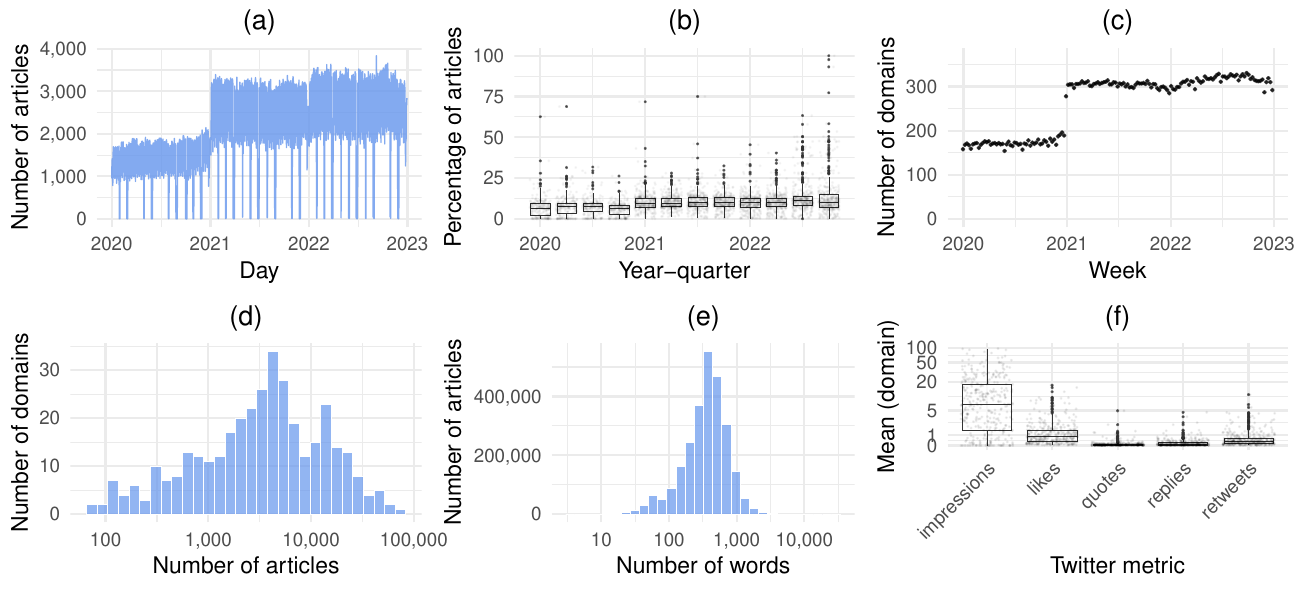}
	\caption{Overview of UKTwitNewsCor dataset distributions for (a) articles per day, (b) percentage of articles per domain per quarter, (c) number of domains per week, (d) articles per domain, (e) words per article, and (f) mean social media metrics across domains.} 
	\label{fig:uktwitnewscor_overview}
\end{figure*}

\subsection{Descriptive Statistics}\label{results}
\begin{figure}[!ht] 
	\includegraphics[width=0.9\linewidth]{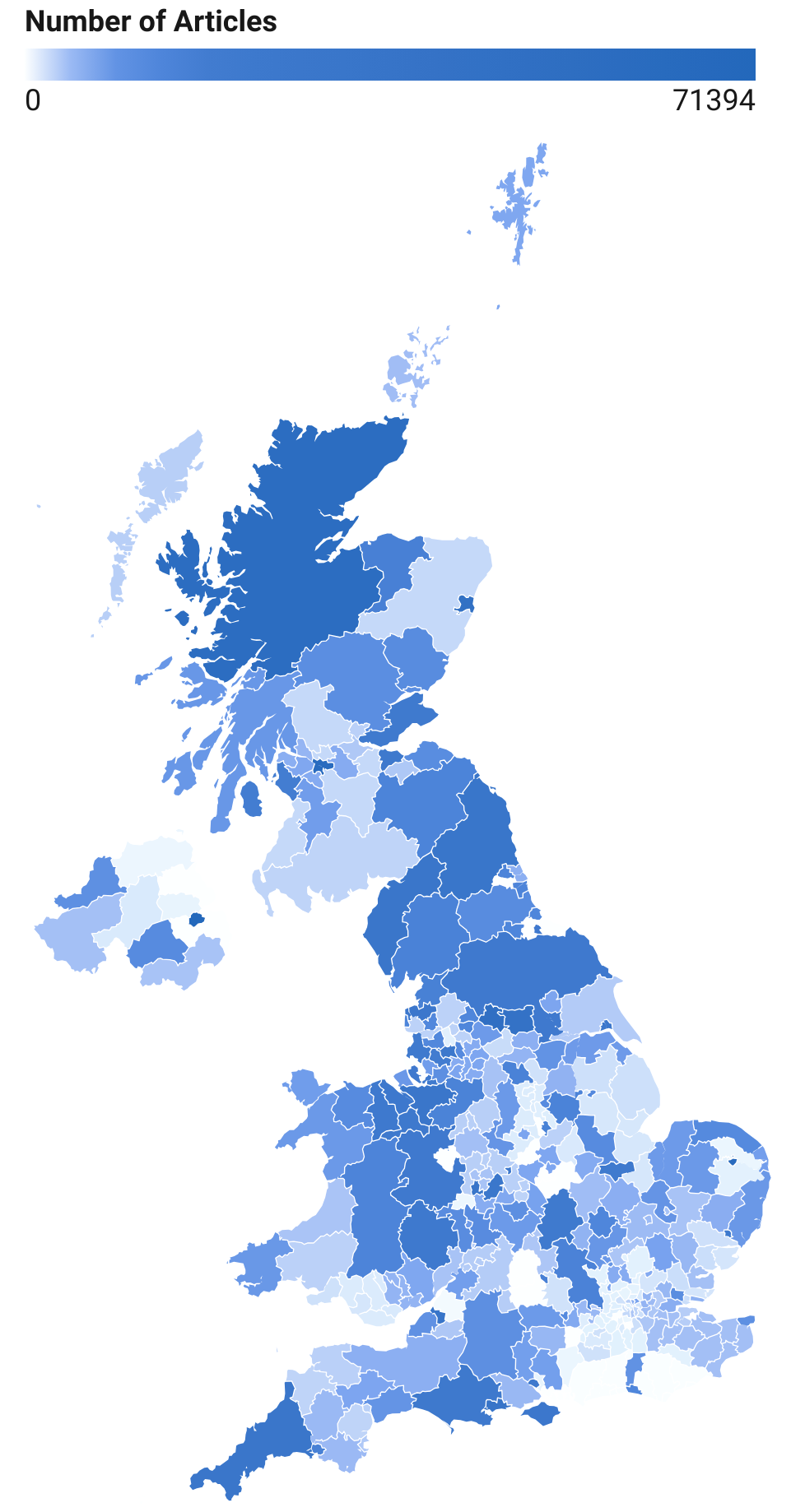}
	\caption{Number of articles per Local Authority District.} 
	\label{fig:article_map}
\end{figure}

The collected articles were sourced from 378 Twitter handles and overall represent 360 domains. This discrepancy emerges from the presence of several Twitter handles linked to the same domain, due to some Twitter accounts reflecting print editions while being consolidated under the same digital site. On average, each domain has 7,041 articles (standard deviation = 9,687) (Table \ref{tab:descriptives}). The distribution of articles by domain resembles a normal distribution, slightly left-skewed (plot (d), Figure \ref{fig:uktwitnewscor_overview}). The total number of articles gradually increases year on year, with a sudden spike from 2020 to 2021 (plot (a), Figure \ref{fig:uktwitnewscor_overview}). We linked this to undetected data retrieval issues for some domains in the first year of the collection (plot (c), Figure \ref{fig:uktwitnewscor_overview}). Nonetheless, the distribution of articles across domains over time remains relatively stable, suggesting even coverage across domains for the 36-months period of the collection (plot (b), Figure \ref{fig:uktwitnewscor_overview}). In terms of engagement, we observe that 470,590 articles (approximately 18.5\% of the total) were reshared at least once. On average, a reshared article is shared 4.75 times (SD = 6.84), with a maximum of 98 reshares for a single article. Additionally, we identified 80,298 articles (around 3.2\%) as being similar to other articles in the dataset. The average number of similar articles per article in this subset is 2.76 (SD = 1.56), with a maximum of 30 similar articles identified for a given article. Across the whole corpus, the average article length is 446 words (SD = 354), with a minimum of 379 and a maximum of 25,577 words. This indicates a wide range of article lengths within the dataset, potentially reflecting the diversity of news content and reporting styles across different domains and publishers. Finally, we calculated the number of articles per Local Authority District (mean = 7180, sd = 9578, median = 4076, min = 0, max = 71394) (Figure \ref{fig:article_map}). To achieve this, we divided the number of articles per domain by the number of districts covered by the domain. This approximation aims to provide a relative measure of news coverage for each Local Authority District, normalising the expected number of articles that a regional domain might produce for each district when covering several.\newline
We found that, while the total number of domains covered in \texttt{UKTwitNewsCor} represents 47\% of all known domains (360 out of 770), we are able to cover 94\% of LADs (337 out of 357 districts - there are 361 LADs in the UK, so four LADs lack domains altogether \cite{bisiani_uncovering_2023}) to some extent. This high coverage percentage can be explained by the tendency of some domains to operate across multiple LADs. When accounting for all domains, including those missing in  \texttt{UKTwitNewsCor}, the mean number of domains across districts is 3.6 (median 3, standard deviation 2.5). The corresponding figure in \texttt{UKTwitNewsCor} is 2.1 (median 2, standard deviation 1.4). This indicates a slight underrepresentation in terms of the total volume of titles per district but suggests that most LADs still have at least one title included in the dataset. The owners coverage is notably lower, with only 39\% of local media owners represented in \texttt{UKTwitNewsCor}. This lower coverage reflects the statistical consequence of a highly skewed market distribution, where the concentration of outlets among a few large media groups, combined with the predominance of single-outlet publishers, creates a sampling bias that systematically reduces the probabilistic representation of media owners in the dataset. The mean number of titles across providers is 3.5 (median 1, standard deviation 15.4), while the corresponding figure in \texttt{UKTwitNewsCor} is 1.6 (median 0, standard deviation 9.9). The median of 0 in UKTwitNewsCor suggests that larger media groups are more likely to be represented, possibly due to their greater digital presence and activity on social media but also due to the greater likelihood of inclusion in the dataset when owning several outlets. 

\begin{table}[t]
\centering
\setlength{\tabcolsep}{1mm}
\begin{tabular}{lrrrrrrrr}
\textbf{Variable} & \textbf{Total} & \textbf{Mean} & \textbf{SD} & \textbf{Min-Max}\\ 
Articles     & 2,534,705 & -     & -    & -  \\
\hspace{0.3cm}\textit{Reshared}  & 470,590  & 4.75      & 6.84     & 2-98 \\
\hspace{0.3cm}\textit{Similar}  & 80,298   & 2.76      & 1.56     & 2-30 \\
Publishers & 87       & 29,135    & 127,135  & 68-873,030 \\
Domains  & 360       & 7,041     & 9,687    & 68-68,694  \\
Words    & 1,130,475,786       & 446     & 354    & 379-25,577 \\
Retweet      &   -  & 0.769     & 5.81     & 0-1999  \\
Reply         &  -  & 0.604     & 5.29     & 0-1986  \\
Likes          &  - & 2.19      & 21.1     & 0-9670  \\
Quote           & - & 0.219     & 3.75     & 0-3517  \\
Impressions     & - & 29.3      & 587.     & 0-262,093 \\
\end{tabular}
\caption{Descriptive Statistics of UKTwitNewsCor Dataset.}
\label{tab:descriptives}
\end{table}

\section{Discussion}
In this section we evaluate the inferential strength of our dataset. The completeness of our seed list of Twitter handles is defined by our approach to Twitter handle collection, which leveraged an automated script searching through the html of each domain homepage. If a handle existed, but the outlet did not display it on the website, it will have escaped our collection. Due to the imminent closure of the Academic Twitter API, we lacked the time to further investigate whether there were additional handles that fell beyond our capturing method. Our twitter data collection covered approximately 50\% of all operative domains within Bisiani and Heravi's (\citeyear{bisiani_uncovering_2023}) directory. We could not find any relevant estimates of how many local media outlets might have a Twitter account, but we guess that a number of smaller, independent outlets might not have had a Twitter account in the period of our study. As such, we expect our Twitter dataset to account for more than half of local media outlets' Twitter accounts active during that period. However, because we used a 2023 local media directory as a seed list for Twitter handles, our tweet dataset is likely to exclude outlets which closed in the period prior to this directory being created, but which were active for some time in the period 2020 to 2022. On the other hand, some outlets were likely launched or consolidated during or after the three year period of our data collection, which might partially explain why we see an increase in the number of articles and domains over time. Because of a lack of comprehensive historical data on outlet closures and launches in the UK \cite{bisiani_uncovering_2023}, we are unable to more granularly provide details about the industry composition at the different points in time covered by our data collection, and our dataset must as such be understood as representing the local media landscape at the time of the data collection (January 2023). While we were unable to further improve our dataset in these respects, we believe its large size and significant coverage make it a useful resource for local media research, particularly in that it allows information about social media metrics to be attached to each article.\newline
Reliance on X (then Twitter) for data collection precludes two foreseen limitations. Firstly, there is an outlet self-selection effect in that not all outlets are on X - we provide data for fewer domains than measured by Moore and Ramsay \shortcite{moore_local_2024}, who listed a total of 630 domains. Nonetheless, we have conducted evaluation checks to empirically verify our data gaps. The additional provision of a complete domain directory, which identifies a further 141 websites relative to Moore and Ramsay, allows for strategic sampling and thorough coverage evaluation. Secondly, there is an article selection effect in that not all articles written might be published on X. In this regard, the article dataset created here is representative of the type of articles that are chosen to be shared on social media across the whole pool of articles by a given outlet. Nonetheless, we are confident that the vast size of the dataset, surpassing any other known article collection for local media outlets in the UK, can provide meaningful inference of the type of news different publishers produce, and an overview of how news might vary across time, geography, and ownership.\newline
By providing coverage metrics at both the district and ownership levels, we are able to  highlight the strengths and limitations of \texttt{UKTwitNewsCor} in representing the full spectrum of local media activity across the UK. Overall, the high coverage of LADs demonstrates that the dataset is geographically comprehensive, though it may under-represent smaller, less active outlets or owners. This is an important consideration for researchers intending to use the dataset to investigate ownership-content diversity or the role of independent publishers in local news ecosystems. With our supplementary datasets on districts and publishers in hand, researchers can devise purposive sampling designs or apply weights to ensure a suitable representative sample of articles is used in their study. While leveraging the Academic Twitter API is no longer feasible due to its deprecation, future UK local media datasets could focus on identifying and including under-represented domains.\newline
In consideration of the closure of the Academic API, our dataset offers a valuable historical archive of local media on the social network platform X. For similar large-scale data collections to occur without incurring in significant costs, future research on local media on social networks will need to adapt by exploring alternative data sources or methods.\newline
The tweet-level performance metrics associated with each news article, including likes, retweets, impressions, and replies, allow researchers to assess not just the popularity of local news stories, but also the dynamics of social media engagement. These metrics can shed light on how stories related to public health, local government, or community events resonate with audiences, and whether certain types of content amplify community discussions or concern. However, researchers should exercise caution when interpreting these metrics, as aggregated engagement data may obscure underlying patterns. Popularity is not synonymous with virality, and engagement levels alone may not capture the full impact of a news story \cite{kormelink_what_2018}. Particularly, reply data can provide deeper insights into audience sentiment and the public’s response to critical issues. Future work could explore the use of reply metadata to better understand which topics spark deeper discussions, or to examine the relationship between engagement levels and content type in more detail.\newline
The inclusion of social media engagement metrics at the article level also opens opportunities to investigate how news outlets with diverse ownership structures adapt their content strategies in response to changes in engagement metrics, such as likes, shares, and retweets. This line of inquiry is particularly relevant for understanding whether higher engagement on specific types of content, such as sensational stories or public interest news, influences editorial priorities or shifts in production. However, caution must be exercised when interpreting these metrics, as they are inherently tied to the dynamics of Twitter (now X). Platform-specific factors, such as algorithmic changes or shifts in policy under new ownership, may distort engagement trends and limit the generalisability of findings to other contexts. Consequently, while the dataset offers valuable insights into platform-mediated behaviours, researchers must exercise caution to ensure robust and contextually grounded interpretations. 
\begin{figure} 
	\includegraphics[width=\linewidth]{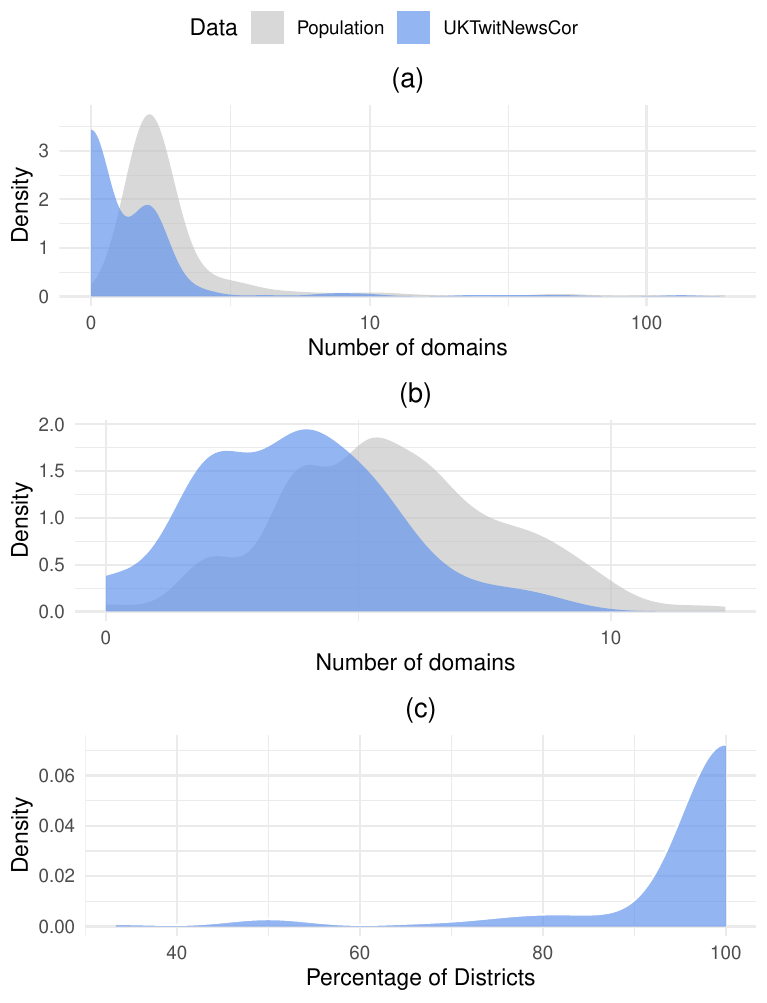}
	\caption{Coverage Evaluation of UKTwitNewsCor dataset with density plots for (a) number of domains by ownership vs population, (b) number of domains by district vs population, and (c) percentage of districts in which a publisher is present in UKTwitNewsCor respective to the ground truth.} 
	\label{fig:coverage}
\end{figure}
\section{Use Cases}
We provide a non-exhaustive overview of potential use cases for the datasets, addressing key research themes and emerging questions in local media studies.
\begin{itemize}
\item \textit{Defining Local Media through Spatial Coverage}
The \texttt{UKTwitNewsCor} dataset enables a content-driven definition of local media by analysing the geographic scope of news coverage. Researchers can delineate "localness" empirically by identifying the regions consistently reported on by specific outlets, addressing a gap previously identified by Hagar \shortcite{hagar_defining_2020}.
\item \textit{Examining Content Repurposing Across Domains}
Researchers can investigate the extent of content repurposing within the local media ecosystem. This analysis sheds light on how large publishers reuse material across domains, potentially prioritising content repurposing over original local reporting \cite{fenton_deregulation_2011, davies_flat_2011}. Patterns of duplication can highlight regional disparities, thematic preferences, and over-reliance on repurposed content.
\item \textit{Impact of Ownership Consolidation on News Topics}
Researchers can examine the effects of ownership consolidation and production centralisation on local relevance of news topics. Analyses of spatial coverage and thematic focus across ownership structures can reveal biases or shifts in reporting priorities, offering insights into the implications of media ownership consolidation for media plurality and democratic discourse \cite{franklin_downgrading_2006}.
\item \textit{Evaluating Local News Provision}
\texttt{UKTwitNewsCor} supports comprehensive analyses of local news provision across the UK. Content-based approaches, inspired by recent international studies \cite{vogler_investigating_2023, khanom_news_2023, madrid-morales_computational_2023}, enable precise identification of underserved communities. Methodologically, the Critical Information Needs (CIN) framework \cite{friedland_review_2012} can be applied to assess information provision over three years.
\item \textit{Analysing Patterns of News Diffusion}
Tweet-level performance metrics facilitate the study of news dissemination on social media. Researchers can evaluate the reach and engagement of stories on topics such as public health or local governance, identifying factors influencing virality and resonance with audiences \cite{cairncross_cairncross_2019}. These insights can inform strategies to amplify the visibility of critical local journalism.
\end{itemize}

\section{Conclusion}
This study introduces \texttt{UKTwitNewsCor}, a novel dataset encompassing over 2.5 million articles from UK online local media. \texttt{UKTwitNewsCor} addresses the "problem of scale" in empirical research of local media outputs by enabling researchers to analyse local journalism with granularity across geography, time, and ownership. Our methodological framework, leveraging digital trace data from Twitter to collect local news articles, demonstrates the viability of using social media as a lens for studying local news ecosystems at scale. This design allows for unique insights into how local news outlets engage audiences and how algorithmic amplification shapes the visibility of public interest news. The longitudinal structure and extensive geographic scope of the dataset allow to examine trends in local journalism, including the impacts of ownership consolidation, regional disparities in news access, and the provision of public interest news provision during a challenging time for the local media sector. The dataset also supports studies of historical events such as the Covid-19 pandemic, Brexit, and the 2019 General Election.\newline
While this work acknowledges the limitations associated with reliance on Twitter to collect data, the inclusion of social media metrics provides a unique opportunity to explore the interplay between algorithmic environments, audience engagement, and editorial control. We anticipate that this dataset will catalyse future studies of digital journalism in the UK, contributing to a more nuanced understanding of the production and availability of local journalism. We hope these findings will advance media and communications theories in the digital age and inform efforts to build a more equitable and resilient local media landscape, particularly as policymakers debate interventions to sustain public interest journalism for all UK communities.

\bibliography{aaai25}

\section*{Paper Checklist}

\begin{enumerate}

\item For most authors...
\begin{enumerate}
    \item  Would answering this research question advance science without violating social contracts, such as violating privacy norms, perpetuating unfair profiling, exacerbating the socio-economic divide, or implying disrespect to societies or cultures?
    \answerYes{Yes, this dataset fills a resource gap in the
study of UK local media without any of the above mentioned negative implications. Our dataset consists of freely available news articles, and if an attempt was made to scrape data from outlets that have paywalls or other pay-to-access services these articles were subsequently screened out as detailed in section "Articles Collection".}
  \item Do your main claims in the abstract and introduction accurately reflect the paper's contributions and scope?
    \answerYes{Yes, the abstract and introduction accurately address the dataset presented in this paper.}
   \item Do you clarify how the proposed methodological approach is appropriate for the claims made? 
    \answerYes{While we do not make any claims beyond describing the dataset, we motivate our approach to data collection in the "Discussion" section.}
   \item Do you clarify what are possible artifacts in the data used, given population-specific distributions?
    \answerNA{NA}
  \item Did you describe the limitations of your work?
    \answerYes{Yes, we describe limitations at length in the "Discussion" section.}
  \item Did you discuss any potential negative societal impacts of your work?
    \answerNo{No, as we do not foresee any potential negative social impacts resulting from this work.}
      \item Did you discuss any potential misuse of your work?
    \answerYes{Yes, we describe in section "Distribution, FAIR Principles, and Use Guidance" the considerations we envision data users will need to apply.}
    \item Did you describe steps taken to prevent or mitigate potential negative outcomes of the research, such as data and model documentation, data anonymization, responsible release, access control, and the reproducibility of findings?
    \answerYes{Yes, any relevant information can be found under section "Distribution, FAIR Principles, and Use Guidance".}
  \item Have you read the ethics review guidelines and ensured that your paper conforms to them?
    \answerYes{Yes, we have read the ethics review guidelines and believe our dataset and this paper conforms to them.}
\end{enumerate}

\item Additionally, if your study involves hypotheses testing...
\begin{enumerate}
  \item Did you clearly state the assumptions underlying all theoretical results?
    \answerNA{NA}
  \item Have you provided justifications for all theoretical results?
    \answerNA{NA}
  \item Did you discuss competing hypotheses or theories that might challenge or complement your theoretical results?
    \answerNA{NA}
  \item Have you considered alternative mechanisms or explanations that might account for the same outcomes observed in your study?
    \answerNA{NA}
  \item Did you address potential biases or limitations in your theoretical framework?
    \answerNA{NA}
  \item Have you related your theoretical results to the existing literature in social science?
    \answerNA{NA}
  \item Did you discuss the implications of your theoretical results for policy, practice, or further research in the social science domain?
    \answerNA{NA}
\end{enumerate}

\item Additionally, if you are including theoretical proofs...
\begin{enumerate}
  \item Did you state the full set of assumptions of all theoretical results?
    \answerNA{NA}
	\item Did you include complete proofs of all theoretical results?
    \answerNA{NA}
\end{enumerate}

\item Additionally, if you ran machine learning experiments...
\begin{enumerate}
  \item Did you include the code, data, and instructions needed to reproduce the main experimental results (either in the supplemental material or as a URL)?
    \answerNA{NA}
  \item Did you specify all the training details (e.g., data splits, hyperparameters, how they were chosen)?
    \answerNA{NA}
     \item Did you report error bars (e.g., with respect to the random seed after running experiments multiple times)?
    \answerNA{NA}
	\item Did you include the total amount of compute and the type of resources used (e.g., type of GPUs, internal cluster, or cloud provider)?
    \answerNA{NA}
     \item Do you justify how the proposed evaluation is sufficient and appropriate to the claims made? 
    \answerNA{NA}
     \item Do you discuss what is ``the cost`` of misclassification and fault (in)tolerance?
    \answerNA{NA}
  
\end{enumerate}

\item Additionally, if you are using existing assets (e.g., code, data, models) or curating/releasing new assets, \textbf{without compromising anonymity}...
\begin{enumerate}
  \item If your work uses existing assets, did you cite the creators?
    \answerYes{Yes, UKTwitNewsCor was developed based on a pre-existing list of digital local media outlets compiled by Bisiani and Heravi (2023). We have properly credited the authors and clearly specified which parts of the dataset are derived from their work. Additionally, we transparently note the inclusion of an extra variable on population size, which was openly provided by the Office for National Statistics.}
  \item Did you mention the license of the assets?
    \answerYes{The licence for our contributed dataset can be found under section "Distribution, FAIR Principles, and Use Guidance". We report the additional assets used are open, and describe their licenses here: the data by Bisiani and Heravi (2023) is supplied under a CC BY V4.0 license. Data from the Office for National Statistics is openly provided under the Open Government License (OGL).}
  \item Did you include any new assets in the supplemental material or as a URL?
    \answerYes{Yes, a URL link to the Harvard Dataverse DOI is provided at the beginning of this paper to lead readers to the data persistent location.}
  \item Did you discuss whether and how consent was obtained from people whose data you're using/curating?
    \answerNA{NA}
  \item Did you discuss whether the data you are using/curating contains personally identifiable information or offensive content?
    \answerNA{NA}
\item If you are curating or releasing new datasets, did you discuss how you intend to make your datasets FAIR?
\answerYes{Yes, this was done in section "Distribution, FAIR Principles, and Use Guidance"}
\item If you are curating or releasing new datasets, did you create a Datasheet for the Dataset? 
\answerNo{No. However, we ensured that the paper submitted here covers the applicable questions by Gebru et al. (2021). Relevant information can be found across section "Methodology" and "Dataset Overview".}
\end{enumerate}

\item Additionally, if you used crowdsourcing or conducted research with human subjects, \textbf{without compromising anonymity}...
\begin{enumerate}
  \item Did you include the full text of instructions given to participants and screenshots?
    \answerNA{NA}
  \item Did you describe any potential participant risks, with mentions of Institutional Review Board (IRB) approvals?
    \answerNA{NA}
  \item Did you include the estimated hourly wage paid to participants and the total amount spent on participant compensation?
    \answerNA{NA}
   \item Did you discuss how data is stored, shared, and deidentified?
    \answerNA{NA}
\end{enumerate}

\end{enumerate}

\section{Appendix}
\subsection{A}
This appendix provides details on the procedure followed to identify duplicated content. Initially, the articles were organised by week to improve computational efficiency. This step was based on observation that duplicate articles were often republished within the same day or a few days. To manage the large dataset, we utilized Locality Sensitive Hashing (LSH), a technique designed to expedite the detection of similar items by hashing (or encoding) articles into compact, manageable signatures \cite{jafari_survey_2021}. Hashing creates a simplified summary of an article, akin to generating a "fingerprint" that captures the core content in a condensed form, making comparison between articles faster and more feasible. We used the MinHash algorithm to create these signatures, which maintains the similarity relationships between articles while significantly reducing the text data size \cite{broder_resemblance_1997}. To measure the similarity between these hashed representations, we applied the Jaccard similarity metric \cite{niwattanakul2013using}. The Jaccard similarity quantifies the degree of overlap between two sets by calculating the proportion of shared elements, with scores ranging from 0 (no similarity) to 1 (identical). For our analysis, articles with a Jaccard similarity score above a threshold of 0.7 were considered highly similar and were grouped into clusters. Each cluster was tagged with a group identifier. This approach efficiently managed the large-scale data, facilitating effective documentation of duplicate and closely related content. 

\subsection{B}
Here we provide the structure and description of the four supplied datasets:
\begin{table*}[t]
\setlength{\tabcolsep}{1mm}
\begin{tabular}{ll}
\textbf{Column Name} & \textbf{Description} \\
unique\_article\_id & A unique identifier for each article \\
article\_text & The full text content of the article \\
domain & The domain name from which the article originates \\
resolved\_url & The fully resolved URL of the article \\
article\_author & The author(s) of the article \\
article\_date & The publication date of the article \\
article\_id & Identifier for reshared articles \\
duplicate\_group & A grouping identifier for articles that are near-duplicates \\
twitter & The Twitter handle that shared the article \\
twitter\_id & The unique identifier of the Twitter handle that shared the article \\
tweet\_url & The original article URL in the tweet \\
tweet\_date & The date the article was shared on Twitter \\
retweet\_count & The number of times the tweet sharing the article was retweeted \\
reply\_count & The number of replies to the tweet sharing the article \\
like\_count & The number of likes the tweet sharing the article received \\
quote\_count & The number of times the tweet sharing the article was quoted \\
impression\_count & The number of impressions the tweet sharing the article received \\
LAD & The Local Authority District associated with the article \\
main\_LAD & The primary or most relevant Local Authority District for the article \\
owner & The owner or publisher of the domain from which the article originates \\
\end{tabular}
\caption{Description of the ARTICLES dataset}
\label{tab:articles_description}
\end{table*}

\begin{table*}[t]
\setlength{\tabcolsep}{1mm}
\begin{tabular}{ll}
\textbf{Column Name} & \textbf{Description} \\
domain & The domain name \\
Owner & The publisher associated with the domain \\
URL & The full URL of the domain \\
LAD & The Local Authority District(s) associated with the domain \\
Main\_LAD & The primary Local Authority District for the domain \\
Twitter & The Twitter handle(s) associated with the domain \\
author\_id & Unique identifier for the author of the articles on the domain \\
n\_articles & The number of articles from this domain \\
like\_stats\_likes\_mean & The mean number of likes for articles from this domain \\
like\_stats\_likes\_median & The median number of likes for articles from this domain \\
like\_stats\_likes\_sd & The standard deviation of the number of likes for articles from this domain \\
retweet\_stats\_retweets\_mean & The mean number of retweets for articles from this domain \\
retweet\_stats\_retweets\_median & The median number of retweets for articles from this domain \\
retweet\_stats\_retweets\_sd & The standard deviation of the number of retweets for articles from this domain \\
impression\_stats\_impressions\_mean & The mean number of impressions for articles from this domain \\
impression\_stats\_impressions\_median & The median number of impressions for articles from this domain \\
impression\_stats\_impressions\_sd & The standard deviation of the number of impressions for articles from this domain \\
reply\_stats\_replies\_mean & The mean number of replies for articles from this domain \\
reply\_stats\_replies\_median & The median number of replies for articles from this domain \\
reply\_stats\_replies\_sd & The standard deviation of the number of replies for articles from this domain \\
quote\_stats\_quotes\_mean & The mean number of quotes for articles from this domain \\
quote\_stats\_quotes\_median & The median number of quotes for articles from this domain \\
quote\_stats\_quotes\_sd & The standard deviation of the number of quotes for articles from this domain \\
\end{tabular}
\caption{Description of the DOMAINS dataset}
\label{tab:domains_description}
\end{table*}

\begin{table*}[t]
\setlength{\tabcolsep}{1mm}
\begin{tabular}{ll}
\textbf{Column Name} & \textbf{Description} \\
LAD & The Local Authority District name \\
LAD\_2021\_pop & The population of the Local Authority District in 2021 \\
Latitude & The latitude of the Local Authority District \\
Longitude & The longitude of the Local Authority District \\
LAD23CD & The code identifier for the Local Authority District \\
n\_domains & The total number of domains associated with this district \\
domains\_owners & The owners of the domains associated with this district \\
n\_present & The number of domains present in the UKTwitNewsCor dataset for this district \\
n\_missing & The number of domains missing from the UKTwitNewsCor dataset for this district \\
percentage\_covered & The percentage of domains covered by the UKTwitNewsCor dataset for this district \\
n\_articles\_lad & The total number of articles from this district in the UKTwitNewsCor dataset \\
articles\_per\_100k & The number of articles per 100,000 population for this district \\
\end{tabular}
\caption{Description of the DISTRICTS dataset}
\label{tab:districts_description}
\end{table*}

\begin{table*}[t]
\setlength{\tabcolsep}{1mm}
\begin{tabular}{ll}
\textbf{Column Name} & \textbf{Description} \\
owner & The publisher \\
domains & The domains owned by this publisher \\
n\_domains & The number of domains owned by this publisher \\
n\_present & The number of domains present in the UKTwitNewsCor dataset for this publisher \\
n\_missing & The number of domains missing from the UKTwitNewsCor dataset for this publisher \\
perc\_present & The percentage of domains present in the UKTwitNewsCor dataset for this publisher \\
lads & The Local Authority Districts covered by this publisher's domains \\
n\_lads & The number of Local Authority Districts covered by this publisher's domains \\
domains\_articles & The list of domains for which there are articles \\
lads\_articles & The total number of articles associated with the Local Authority Districts covered by this publisher \\
n\_lads\_articles & The number of Local Authority Districts for which there are articles from this publisher \\
n\_articles & The total number of articles from this publisher in the UKTwitNewsCor dataset \\
mean\_articles & The mean number of articles per domain for this publisher \\
median\_articles & The median number of articles per domain for this publisher \\
sd\_articles & The standard deviation of the number of articles per domain for this publisher \\
\end{tabular}
\caption{Description of the PUBLISHERS dataset}
\label{tab:publishers_description}
\end{table*}

\end{document}